# Influence of sintering temperature on resistivity, magnetoresistance and thermopower of $La_{0.67}Ca_{0.33}MnO_3$

G Venkataiah[1,2], YK Lakshmi[1] and PV Reddy*[1]

Address: [1]Department of Physics, Osmania University, Hyderabad 500 007, India and [2]Department of Physics, National Cheng Kung University, Tainan 701, Taiwan, Republic of China ROC

Email: G Venkataiah - gvenkataiah@gmail.com; YK Lakshmi - kalyaniraj@osmania.ac.in; PV Reddy* - pvenugopalreddy@yahoo.com

* Corresponding author

## Abstract

A systematic investigation of $La_{0.67}Ca_{0.33}MnO_3$ manganites has been undertaken, mainly to understand the influence of varying crystallite size (nano meter range) on electrical resistivity, magnetic susceptibility and thermoelectric power. The materials were prepared by the sol-gel method of sintering at four different temperatures between 800–1100°C. The samples were characterized by X-ray diffraction and data were analyzed using Rietveld refinement. The metal-insulator transition temperatures ($T_P$) are found to increase with increasing sintering temperatures, while the magnetic transition temperatures ($T_C$) decrease.

The electrical resistivity and thermoelectric power data at low temperatures ($T < T_P$) have been analyzed by considering various scattering phenomena, while the high temperature ($T > T_P$) data were analyzed with Mott's small polaron hopping conduction mechanisms.

## 1. Introduction

In recent years, nanocrystalline materials have attracted the attention of the scientific community because of the rich Physics involved as well as their potential use in device applications [1-6]. The presence of a large amount of grain boundaries and/or the broad distribution of interatomic spacings in the grain boundaries give rise to the unusual properties of nanocrystalline materials when compared to conventional polycrystals or single crystals with the same chemical composition. Magnetic nanoparticles with smaller grain sizes exhibit richer electronic and magnetic properties arising from structural and magnetic disorders in the grain surfaces. Recently, modification of the properties of nanosized perovskites has aroused much interest [7-10]. Doped manganites with strongly correlated electrons exhibit fascinating properties originating from the strong inter-play between charge, spin, orbital, and lattice degrees of freedom. Doped manganites $R_{1-x}A_xMnO_3$ (R = rare earth trivalent cation and A = divalent alkaline earth cation) undergo a paramagnetic insulator (PMI) to ferromagnetic metal (FMM) transition and exhibit colossal magnetoresistance (CMR) phenomena in the vicinity of their transition temperature. The observed correlation between the metallicity and ferromagnetism in manganites has been explained within the framework of the double exchange (DE) mechanism, which describes the electronic hopping between neighboring Mn ($e_g$) orbitals. However, the DE mechanism cannot explain the entire observed phase diagram. In addition to DE, the polaron effect due to a very strong electron phonon coupling coming from the Jahn-Teller (JT) lattice distortion of the $Mn^{3+}$ is very helpful in explaining the resistivity and magnetoresistance of these compounds [11]. Among various aspects of transport studies in these materials, thermopower has attracted much attention, because it is a very sensitive physical property and depends on the nature of charge carriers and their interaction with spins. Although a lot of work on nanocrystalline manganites, based on their resistivity and magnetoresistance, is available in the literature, data on thermopower of these materials are scanty. In view of these facts, a systematic investigation of electrical resistivity, magnetoresistance and thermopower studies of $La_{0.67}Ca_{0.33}MnO_3$ manganites have been carried out and the results of such an investigation are presented here.

## 2. Experimental details

Nanosized polycrystalline samples with compositional formula, $La_{0.67}Ca_{0.33}MnO_3$ were synthesized by the sol-gel route, taking corresponding metal nitrates as starting materials in a stoichiometric ratio. Later, these solutions were converted into citrates and the $p^H$ was adjusted between 6.5 and 7. After getting a sol on slow evaporation, a gelating reagent ethylene glycol was added and heated on a hot plate between 160 and 180°C to get a gel. Finally, the resulting powder was pressed into circular pellets, which were sintered in air atmosphere for 4 hrs at 800, 900, 1000 and 1100°C to produce samples of different particle sizes. All the samples were characterized by X-ray diffraction (XRD) (Philips xpert diffractometer). The XRD data were analyzed using Rietveld refinement technique [12] and the average crystallite sizes were estimated using peak broadening. The electrical resistivity and magnetoresistance (MR) measurements were performed using JANI'S 'supervaritemp' cryostat in an applied magnetic field of 0,1,3,5 and 7T over a temperature range 77–300 K and magnetic transition temperatures ($T_C$) were determined by measuring AC susceptibility ($\chi$) in the temperature range 77–300 K by using the mutual inductance principle. Finally, a dynamic two probe differential method was employed for the measurement of the thermopower [13]. The samples were attached with silver paint between two copper electrodes with an adjustable temperature gradient and were monitored using a copper-constantan thermocouple. The assembly was placed in a closed liquid nitrogen cryostat and nitrogen atmosphere was used as an exchange, to maintain uniform temperature and avoid condensation of moisture. The data were collected in heating mode. The measured thermopower data were corrected by subtracting thermopower values of copper, so as to obtain the absolute thermopower values of the samples.

## 3. Results and discussion

### 3.1. X-ray diffraction

The XRD pattern of $La_{0.67}Ca_{0.33}MnO_3$ samples, sintered at different temperatures are shown in Fig. 1. The XRD data have been analyzed with the Rietveld refinement technique and the materials were found to crystallize in Pbnm space group. It is also clear from the analysis that the samples are single phase with no detectable impurity. The Mn-O-Mn bond angle and Mn-O bond length obtained from the Rietveld refinement are presented in Table 1. The average particle sizes of the materials were estimated using peak broadening [10] through the Scherrer formula, <S> = Kλ/βcosθ, where <S> average crysttallite size in Å, K is a constant (shape factor; 0.89), λ is the Cu Kα wavelength and β is the corrected full-width-half-maxima of XRD peaks of the sample. $SiO_2$ was used to correct the intrinsic width associated with the equipment. The calculated average crystallite size values are given in Table 1. It is clear from the table that the crystallite sizes are found to increase with increasing sintering temperature.

### 3.2. Magnetic and electrical

AC susceptibility measurements of all the samples have been carried out as a function of temperature (Fig. 2) and based on these results, the ferro to paramagnetic transition temperatures ($T_C$) were obtained and are given in Table 2. It is clear from the table that $T_C$ values decrease with increasing sintering temperature thereby indicating that $T_C$ is decreasing with increasing particle size. The observed behavior may be explained following the work by Dutta et al. [14]; According to these authors, the magnetic and transport properties of the perovskite manganites are strongly coupled and are very sensitive to Mn-O-Mn bond angle and Mn-O bond length. It has been found that a decrease in magnetization and an increase in resistivity occur as we decrease the particle size, due to broken Mn-O-Mn bonds at the surface of smaller particles that hamper exchange interaction and degrade connectivity for electron conduction, but in this case it seems that the spin interaction increases and the connectivity improves as we decrease the particle size. The decrease in particle size results increase in Mn-O-Mn bond angle and decrease in Mn-O bond length. Therefore magnetization increases and $T_C$ enhances with decreasing particle size. It is also clearly evident from Table 1 that the bond angles are found to be almost constant, while the bond lengths decrease continuously with decreasing particle size of the material. In view of these arguments and explanations, it is reasonable to understand that there is a continuous decrease in the values of $T_C$ with a continuous increasing sintering temperature.

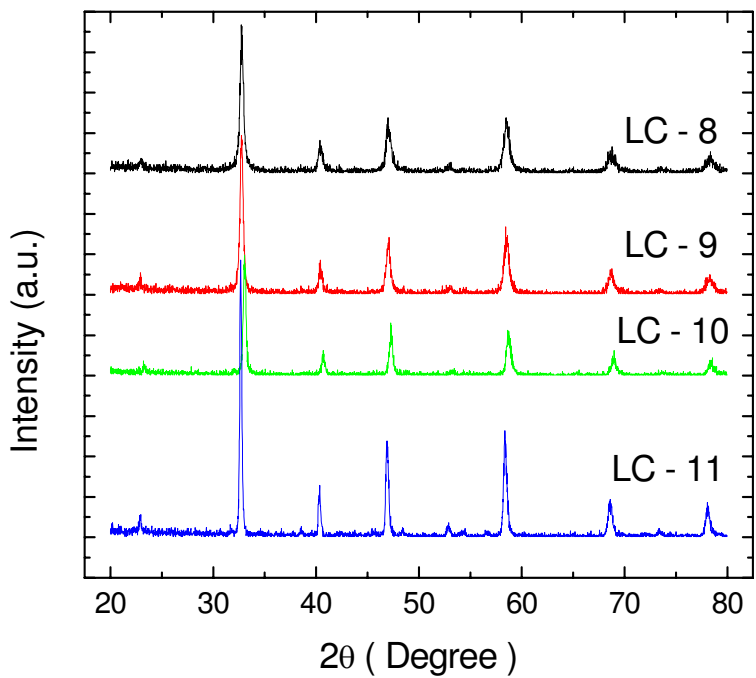

**Figure 1**
X-ray diffraction patterns of $La_{0.67}Ca_{0.33}MnO_3$ manganites.

**Table 1: Experimental data of LCMO manganites**

| Compositional formula | Sample Code | Sintering temp. (°C) | <S> (nm) | a(Å) | b(Å) | c(Å) | Mn-O-Mn (Degree) | Mn-O (Å) |
|---|---|---|---|---|---|---|---|---|
| $La_{0.67}Ca_{0.33}MnO_3$ | LC-8 | 800 | 20 | 5.466(3) | 5.449(2) | 7.727(5) | 148.5 | 1.82205 |
| $La_{0.67}Ca_{0.33}MnO_3$ | LC-9 | 900 | 30 | 5.468(3) | 5.446(2) | 7.742(3) | 148.4 | 1.82345 |
| $La_{0.67}Ca_{0.33}MnO_3$ | LC-10 | 1000 | 35 | 5.475(1) | 5.462(1) | 7.732(3) | 148.5 | 1.82430 |
| $La_{0.67}Ca_{0.33}MnO_3$ | LC-11 | 1100 | 40 | 5.480(2) | 5.462(2) | 7.741(6) | 148.5 | 1.82550 |

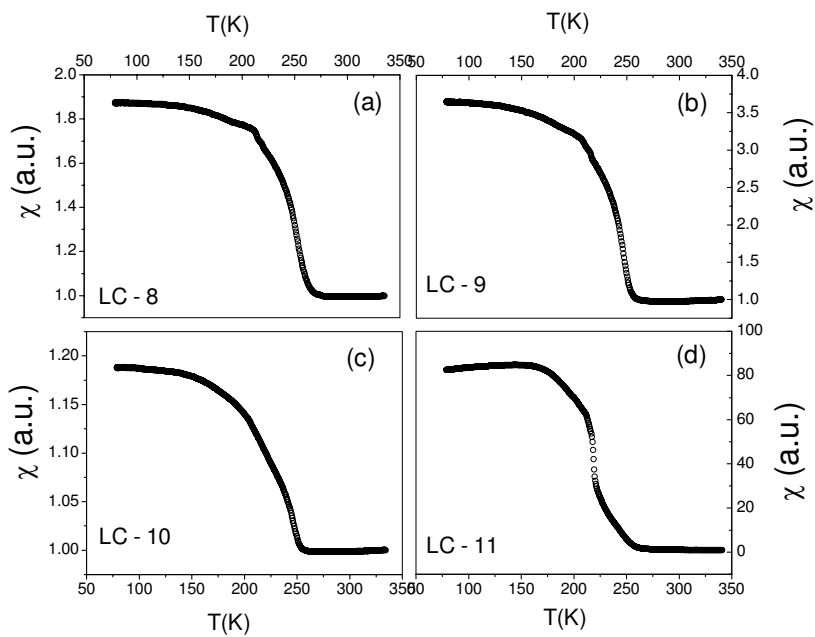

**Figure 2**
Temperature dependence of AC susceptibility of $La_{0.67}Ca_{0.33}MnO_3$ samples.

**Table 2: Electrical and magnetic data of LCMO manganites.**

| Sample Code | $T_C$ (K) | $T_P$ (K) | $T_S$ (K) | $\Delta T = T_C - T_P$ (K) | $\rho_{Peak}$ ($\Omega$cm) | MR% (7T) |
|---|---|---|---|---|---|---|
| LC-8  | 253 | 145 | 229 | 108 | 969 | 63 |
| LC-9  | 248 | 160 | 236 | 88  | 429 | 71 |
| LC-10 | 246 | 180 | 257 | 66  | 176 | 70 |
| LC-11 | 219 | 195 | 266 | 24  | 112 | -  |

The metal insulator transition temperature ($T_P$) and peak resistivity ($\rho_{Peak}$) obtained from electrical resistivity measurements are given in Table 2. It is interesting to note that $T_P$ values are found to increase with increasing sintering temperature from 145 to 195 K, while $\rho_{Peak}$ values are decreasing. In view of these observations, one can understand that there is a tremendous influence of crystallite size on various properties mentioned, and that the observed behavior may be explained on the basis of a qualitative model. According to this model, it has been assumed that in the case of sol-gel prepared samples of the present investigation, when the grain size of the material is decreased, a non-magnetic surface layer having nanocrystalline size would be created around the grain. This may increase the residual resistivity of the material, which in turn decreases the density of Ferromagnetic Metallic (FMM) particles. Therefore, lowering of $T_P$ and enhancement of electrical resistivity at a given temperature ($\rho_T$) are expected. Finally, the shifting of $T_P$ towards the low temperature region could be due to the loss of long-range ferromagnetic order in the sample [15].

It can also be seen that a difference between $T_C$ and $T_P$ is observed and it is has been calculated $\Delta T$ ($T_C \sim T_P$) for each material and given in Table 2. The $\Delta T$ values are found to decrease from a large value of 108 to 24 K as the sintering temperature increases continuously and in fact, a similar difference between $T_C$ and $T_P$ has been reported earlier [14]. The observed difference between the two transition temperatures may be explained as outlined here. It is well known that two contributions are responsible for the transport properties among CMR materials. One of them is intrinsic and might have originated from the double exchange (DE) interaction between the neighboring Mn ions, while the other one is extrinsic and is due to spin-polarized tunneling between ferromagnetic grains through an insulating grain boundary (GB) barrier. According to DE model, the metal-insulator transition always occurs in the vicinity of $T_C$. However, in the case of granular samples with a large number of GBs, the influence of interfaces and boundaries should be taken into account. Further, as the GB is similar to the amorphous state, the magnetic configuration on the grain surface is more disordered than in the core. In such a situation, the occurrence of anti-ferromagnetic insulating regions on the grain boundary may not modify the magnetic transition temperature, $T_C$. However, the phenomenon may influence the metal-insulator (electrical) transition, $T_P$ thereby shifting it to lower temperatures [16].

### *3.3. Magnetoresistance*

Magnetoresistance measurements were carried out in the presence of different magnetic fields viz; 1, 3, 5 and 7 T. A typical plot of variation of electrical resistivity with temperature in the case of LC-9 at different field runs is shown in Fig 3. It can be seen from the figure that the resistivity is found to decrease with increasing magnetic field and that $T_P$ shifts towards higher temperatures and that as a matter of fact, $T_P$ values are found to change from 160–185 K when the field changes from 0–7 T. This may be due to the fact that the applied magnetic field induces delocalization of charge carriers, which in turn might suppress the resistivity and also cause local ordering of the magnetic spins. Due to this ordering, the ferromagnetic metallic (FMM) state may suppress the paramagnetic insulating (PMI) regime. As a result, the conduction electrons ($e^1_g$) are completely polarized inside the magnetic domains and are easily transferred between the pairs of $Mn^{3+}$ ($t^3_{2g}$ $e^1_g$: S = 2) and $Mn^{4+}$($t^3_{2g} e^o_g$: S = 3/2) via oxygen and hence the peak temperature ($T_P$) shifts to high temperature side with application of magnetic field [17]. In fact a similar explanation was given earlier [18].

Further, the percentage of MR of all the materials (except LC-11) of the present investigation has been calculated in the vicinity of $T_C$ by using the well-known relation,

$$MR\% = \{[\rho_{(0)} - \rho_{(H)}]/\rho_{(0)}\} \times 100 \qquad (1)$$

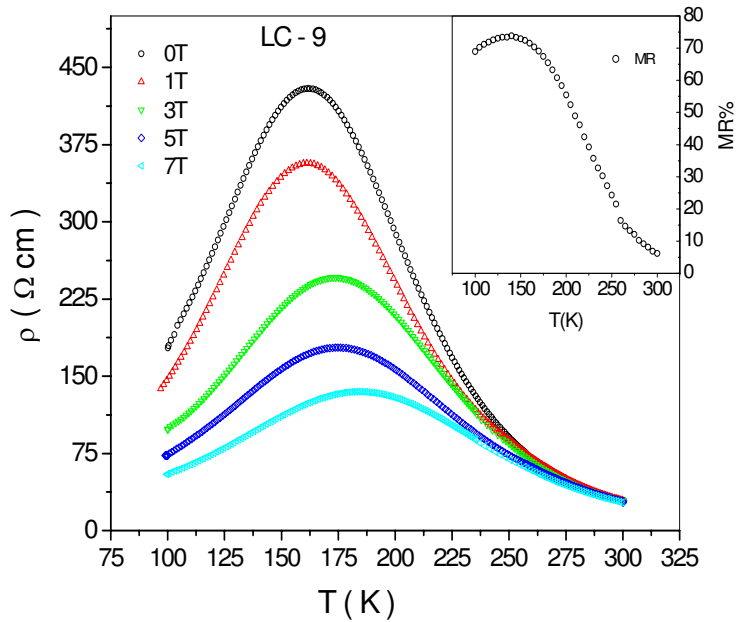

**Figure 3**
Resistivity versus absolute temperature of LC-9, at different magnetic fields.

The calculated MR values are found to remain constant (Table 2) with increasing sintering temperature. The variation percentage of MR of LC-9 with temperature is also shown in the inset of the Fig. 3 and one may note from the figure that the percentage of MR values is found to be very high in the vicinity of $T_C$.

The electrical resistivity data ($T<T_P$) of samples of the present investigation are found to fit well with the equation

$$\rho = \rho_0 + \rho_{2.5} T^{2.5} \qquad (2)$$

Here the term $\rho_0$ represents the resistivity due to grain/domain boundary effects, while the second term, $\rho_{2.5}T^{2.5}$ represents the resistivity due to electron-magnon scattering process in ferromagnetic phase [17].

The conduction mechanism of manganites at high temperatures ($T > T_P$) is explained by adiabatic small polaron hopping mechanism and the equation is given by

$$\rho = \rho_\alpha T \exp(E_P/k_B T) \qquad (3)$$

where $\rho_\alpha$ is residual resistivity, while $E_P$ is the activation energy, $\rho_\alpha = 2k_B/3ne^2a^2\nu$, here $k_B$ is Boltzmann's constant, $e$ is the electronic charge, $n$ is the number of charge carriers, $a$ is site – to – site hopping distance and $\nu$ is the longitudinal optical phonon frequency. The activation energy values, calculated from the best-fit parameters are given in Table 3.

### 3.4. Thermoelectric power

Thermoelectric power (TEP) measurements were carried out over a temperature range 77–300 K and Seebeck coefficient (S) values have been computed at different temperatures. The variation of S with temperature for all the samples is shown in Fig 4 and it can be seen that the magnitude of S decreases with increasing sintering temperature. It is also clear from the figure that in the case of LC-8, LC-9 and LC-10 samples the sign of S is positive through out the temperature range of investigation, while in the case of LC-11, thermopower changes from a positive to a negative value. It means that the samples with lower particle size (20, 30 and 35 nm) exhibit a positive S value, while those with higher crystallite size (45 nm) exhibit both negative and positive values. The observed behaviour may be explained as follows. The positive sign exhibited by LC-8, LC-9 and LC-10 materials might be attributed to holes which are excited from the valence band (VB) into the impurity band, while in the case of LC-11 sample, the change in sign may be attributed to the orbital degeneracy of the $e_g$ band. According to this model, the orbital degree of freedom of the $e_g$ band may play an important role. It is well known that $e_g$ band consists of degenerate 3$d$ orbital (i.e., $d_{3z^2-r^2}$ and $d_{x^2-y^2}$) and may split into upper and lower bands by an order of $J_H$ [19]. If the lower (spin-up) band, splits further into two bands in the FM state, then the lowest band is filled, the dopants may introduce holes thereby showing positive S values while the lowest band is empty; it shows negative (electron-like) values. Therefore, one may conclude that there may be a possibility of changing sign from positive to negative with varying particle size [20].

**Table 3: Best fit parameters obtained from of thermopower and electrical resistivity of LCMO manganites.**

| Sample Code | $S_0$ ($\mu VK^{-1}$) | $S_{3/2}$ ($\mu VK^{-5/2}$) | $S_4$ ($\mu VK^{-5}$) | $E_P$ (meV) | $E_S$ (meV) | $W_H = E_P - E_S$ (meV) | $\alpha'$ |
|---|---|---|---|---|---|---|---|
| LC-8  | -3.5850 | 0.0084 | $-3.0890 \times 10^{-9}$ | 160.31 | 26.368 | 133.942 | -0.5994 |
| LC-9  | -5.4796 | 0.0105 | $-1.7713 \times 10^{-8}$ | 157.94 | 22.606 | 135.334 | -0.4768 |
| LC-10 | -6.9553 | 0.0124 | $-2.2135 \times 10^{-8}$ | 155.09 | 17.731 | 137.359 | -0.2747 |
| LC-11 | -2.4868 | 0.0544 | $-1.1094 \times 10^{-8}$ | 143.43 | 2.859  | 140.571 | -0.2410 |

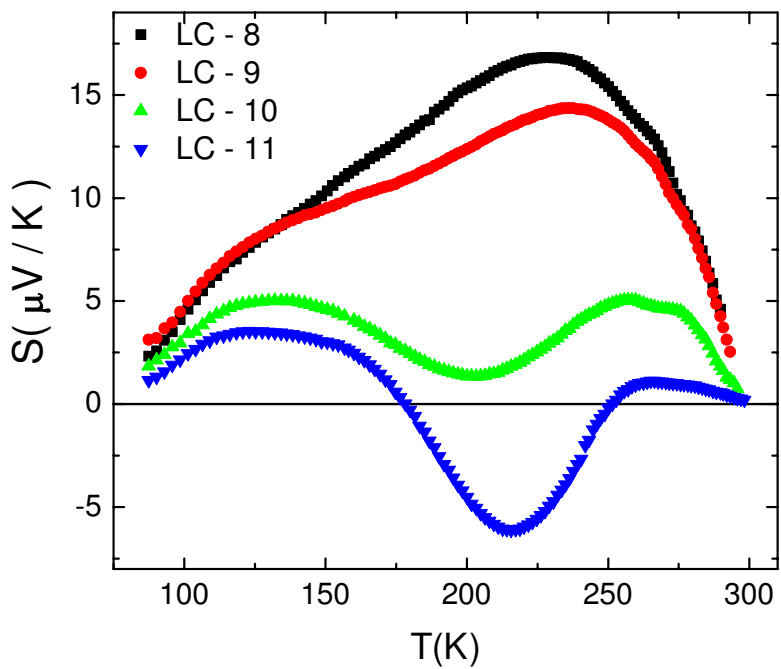

**Figure 4**
Variation of S with T of LCMO manganites.

Close observation of S versus T plots shows that as the temperature decreases from 300 to 77 K, the values of S increases thereby attaining a maximum value in the vicinity of magnetic transition temperature ($T_C$) designated as $T_S$. The obtained $T_S$ values from S vs. T plots are included in the Table 2. Further, $T_S$ values and the magnitude of S at $T_S$ increase with increasing particle size. In the ferromagnetic metallic part ($T < T_P$), a broad peak is found to develop and increase with increased particle size. In fact, similar behaviour was reported earlier in Pr-based manganites [21,22]. The observed broad peak may be explained on the basis of the spin-wave theory. According to this theory, in ferromagnets and antiferromagnets, electrons are scattered by spin waves, giving rise to the electron-magnon scattering effect. In a manner similar to the scattering of phonons resulting in phonon drag effects, the electron-magnon interaction also produces a magnon drag effect. As the magnon drag effect is approximately proportional to the magnon specific heat, one may expect the variation of S with temperature as $T^{3/2}$ for ferromagnetic materials [22]. In view of these arguments, one may conclude that the observed broad peaks may be attributed to the magnon drag effect which increases with increase in particle size.

*3.4.1. Low Temperature ($T < T_P$) behaviour*

Similarly to the electrical conduction, several factors, namely impurity, complicated band structure, electron-electron, magnon scattering, etc, affect the TEP data at low temperatures ($T < T_P$). Therefore, the ferromagnetic metallic part of thermopower data was fitted to an equation:

$$S = S_0 + S_{3/2}T^{3/2} + S_4T^4 \qquad (4)$$

where $S_0$ is a constant that accounts for the low temperature TEP data, while the $S_{3/2}$ term arises from the electron magnon scattering contribution and the origin of the $S_4$ term, which is dominant in the high temperature region especially near the transition temperature, is still not clear.

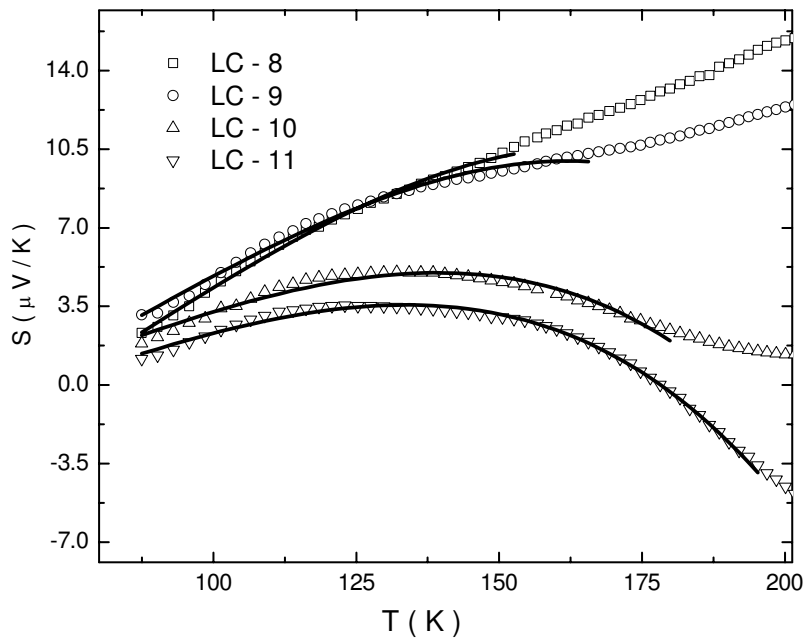

**Figure 5**
**Variation of S with T of LCMO manganites at ($T<T_P$)**. The solid line gives the best fit to the equation $S = S_0 + S_{3/2}T^{3/2} + S_4T^4$.

The best-fit parameters obtained from equation (4) are given in Table 3 and corresponding plots are shown in Fig. 5. It is clear from the table that the $S_0$ values increase with increasing particle size for first three samples (LC-8, LC-9 and LC-10) and decrease in the case of the LC-11 sample, while the $S_{3/2}$ values increase continuously. The $S_4$ values do not vary systematically with varying particle size of the materials. The increase in $S_{3/2}$ values with increasing particle size shows a continuous increase in ferromagnetic magnon strength in these materials. Similarly, $S_0$ represents the temperature independent thermopower.

### 3.4.2. High temperature ($T > T_P$) behaviour

There is a strong experimental evidence for the presence of small polarons at high temperatures in the case of manganites [23]. In fact, pair-density function (PDF) analysis [24,25] of powder neutron scattering data of manganites clearly indicates that the doped holes are likely to be local-ized within one octahedron as a $Mn^{4+}$ ion, forming a single-site polaron (small polarons). These small polarons in the paramagnetic insulating region are in good agreement with the theoretical prediction of entropic localizations. Therefore, it has been concluded that the small polarons might be responsible for the high temperature ($T > T_P$) paramagnetic conduction process in manganites. In view of this, the thermopower data of the present investigation have been analyzed using small polaron hopping model described by the following equation [26],

$$S = k_B/e \ [E_S/k_BT+\alpha'] \quad (5)$$

where, $k_B$ is Boltzmann's constant, e is electron's charge, $E_S$ is the activation energy obtained from TEP data and the $\alpha'$ is a sample dependent constant, which is associated with the spin and the mixing entropy. In addition, $\alpha' < 1$ suggests the conduction mechanism is due to small polarons, while $\alpha' > 2$ represents the conduction may be due to large polarons [27].

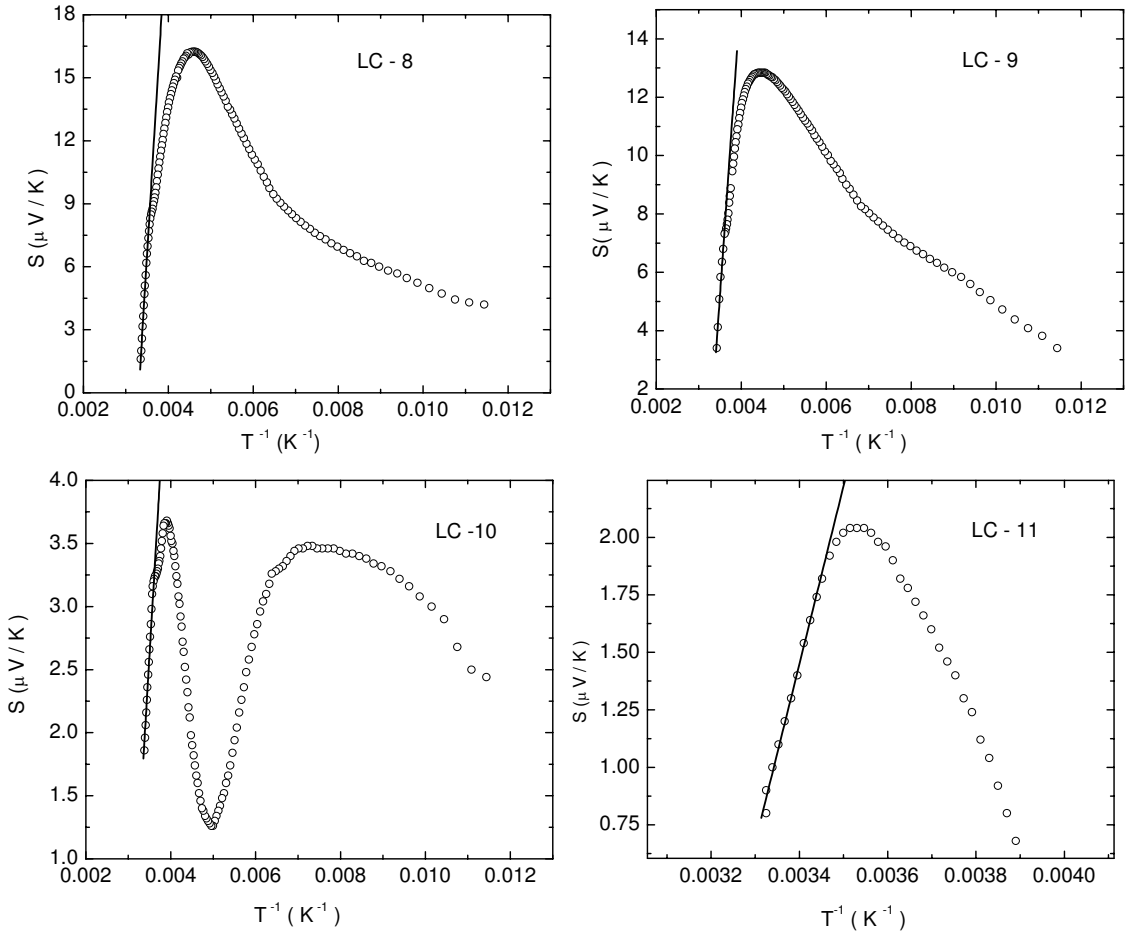

**Figure 6**
**Plots of S versus $T^{-1}$ of LCMO manganites**. The solid line gives the best fit of the equation $S = K_B/e(E_S/K_B T + \alpha')$.

The high temperature TEP data are found to fit well with equation (5) and from the fitting parameters, the $E_S$ and $\alpha'$ values are obtained and given in Table 3. Plots of S versus 1/T of all the samples are shown in Fig. 6. The calculated values show that both the activation energies ($E_P$ and $E_S$) decrease with increasing sintering temperature. Moreover, the magnitude of $E_S$ is much smaller than that of $E_P$, which is a characteristic property of small polaron conduction [27]. The difference between the activation energies, measured from resistivity and TEP studies is the polaron hopping energy $W_H = E_P-E_S$. The decrease in the $E_P$, $E_S$ and $W_H$ may be explained as follows: It is known that with increasing grain size, the interconnectivity between grains increases, which in turn enhances the possibility of conduction electron to hop the neighboring sites [28], thereby conduction bandwidth increases and as a result the value of $E_P$, $E_S$ and $W_H$ values decreases. Therefore one may conclude that the conduction bandwidth may be tuned by varying the particle size of the material.